# An Improved Simulation Model for Pedestrian Crowd Evacuation

**Danial A. Muhammed** [1], **Tarik A. Rashid** [2,*], **Abeer Alsadoon** [3], **Nebojsa Bacanin** [4], **Polla Fattah** [5], **Mokhtar Mohammadi** [6] **and Indradip Banerjee** [7]

[1] Computer Department, College of Science, University of Sulaimani, Sulaymaniyah, 46001, KRG, Iraq; danial.muhammed@univsul.edu.iq

[2] Computer Science and Engineering Department, University of Kurdistan Hewler, Erbil, 44001, KRG, Iraq

[3] Department of Information Technology, School of Computing and Mathematics, Charles Sturt University, Sydney, Bathurst NSW 2795, Australia; alsadoon.abeer@gmail.com

[4] Faculty of Informatics and Computing, Vice-dean, Singidunum University, Belgrade, 11000, Serbia; nbacanin@singidunum.ac.rs

[5] Software and Informatics Engineering Dept, Engineering College, Salahaddin University-Erbil, Erbil, 44001, KRG, Iraq; polla.fattah@su.edu.krd

[6] Department of Information Technology, Lebanese French University, Erbil, 44001, KRG, Iraq; mokhtar.mohammadi1@gmail.com

[7] Department of Computer Science and Engineering, University Institute of Technology, The University of Burdwan, Burdwan, West Bengal, 713101, India; ibanerjee2001@gmail.com

**\*** Correspondence: tarik.ahmed@ukh.edu.krd; Tel.: +964-0750-109-42-33

**Abstract:** This paper works on one of the most recent pedestrian crowd evacuation models, i.e., "a simulation model for pedestrian crowd evacuation based on various AI techniques", developed in late 2019. This study adds a new feature to the developed model by proposing a new method and integrating it with the model. This method enables the developed model to find a more appropriate evacuation area design, among others regarding safety due to selecting the best exit door location among many suggested locations. This method is completely dependent on the selected model's output, i.e., the evacuation time for each individual within the evacuation process. The new method finds an average of the evacuees' evacuation times of each exit door location; then, based on the average evacuation time, it decides which exit door location would be the best exit door to be used for evacuation by the evacuees. To validate the method, various designs for the evacuation area with various written scenarios were used. The results showed that the model with this new method could predict a proper exit door location among many suggested locations. Lastly, from the results of this research using the integration of this newly proposed method, a new capability for the selected model in terms of safety allowed the right decision in selecting the finest design for the evacuation area among other designs.

## 1. Introduction

Currently, population size is intensely increasing, demand for space is inevitable, and new styles of buildings are built extensively due to the quick advancement in economics and its continuity [1,2]. Various structures of these buildings affect the duration of evacuation within the evacuation process [3]. Therefore, considering an operative evacuation system for these buildings is crucial when an emergency state occurs, such as terrorist threats, bombs, fires, and venomous gas [4]. All parties involved, such as residents, governments, and designers, face a problem when an emergency evacuation occurs inside these buildings [2]. Researchers are dependent on modeling to define the communication's rules and conditions between the environment and evacuees when there are deficiencies in the evacuation's realistic data [5]. Accordingly, crowd simulation allows dealing with an emergency, and it is precise, convenient, and supportive [6,7]. In the last two decades, regarding the limitations of involving homogeneous people in various simulation models [8], there were some simulation models proposed for evacuation, such as Simulex [9], BGRAF [10], and Exodus [11]. In 2019, one of the most recent models for simulating pedestrian crowd evacuation using different AI methods was built, which incorporates homogeneous people to simulate the pedestrian evacuation crowd [12]. However, this new model has a limitation that does not allow the best exit door location to be specified for the evacuation area according to the evacuee's efficiency. Moreover, it cannot choose the finest design among several existing designs for the evacuation area.

The main aims of this research are as follows (1) focus on the methodology of this new model and to try and find a method to determine the best exit location for the evacuation area and the best evacuation area design from a safety perspective, and (2) design and implement the method and integrate it with the existing model. Evacuation is a commonly researched field that remains an issue among scientists. There were various research papers on the evacuation process, which tackled different conditions.

The main goal of this paper was to address the increasing demands of using pedestrian crowd simulation models. For safety purposes, governments and architects aim to design buildings properly. Consequently, various evacuation methods appear, such as protective, preventive, rescue, and reconstructive evacuations [13]. The evacuation problem is not solely the physical movement of evacuees; it is multifaceted and related to the physical and social circumstances, such as the high possibility for hazard, great stage of pressure, and inadequate data. These circumstances illustrate robust communication among environment, danger, egress process, population demographics, and participant behavior [14]. Evacuees' communication in a building environment can influence the evacuation system. Therefore, the objective of this work was to offer a methodical technique based on a crowd simulation model to indicate the best exit door location and create a design with more safety. Hence, this paper adds a new environmental ability to the most recent pedestrian evacuation crowd simulation model developed in late 2019. The ability involves determining the best exit door location for evacuation, based on the model's evacuation process results and selecting a more appropriate design for evacuation.

This paper is organized as follows: Section 2 presents a literature review. Section 3 describes the research method of the selected developed model and methodology of this study. Section 4 shows the proposed method with the ability to determine the best exit location for evacuation within an area and presents the simulation results describing the selection of the best exit door locations and indicating a suitable design from the evacuation perspective. Lastly, Section 5 provides the final clarifications and recommends some information for future research work.

## 2. Literature Work

This section reviews several evacuation crowd models that considered the environment, speed, and behavior.

In 2016, S. Nirajan et al., using the collected responses for a questionnaire review of 1127 travelers, constructed and theoretically and mathematically proved a model allowing directors of the train station to find a suitable approach to deal with an emergency via an emergency controller while considering and assessing the locations of emergency exit signs during normal and emergencies in a train station [15]. In 2018, C. Shuchao, et al. offered an extended multi-grid model to examine evacuation within a room with two exit doors under the fire condition. The proposed model could guess the evacuees' movement, exit choice, and act as a guider by providing recommendations to the evacuees when a fire emergency exists [16]. In 2018, Kontou et al. used cellular automata (CA) parallel computing tools in developing a model of crowd evacuation, then within the area of evacuation are used to mimic and assess different appearances and manners of the individuals included disabling. To conduct the simulation process, a secondary school in the region of Xanthi was selected, which included disabling children. The school's safety training was well-ordered with observing and existing earthquake. The evacuation time was recorded entirely. Finally, the realistic data validated the suggested model, and there was an expediency implication to the particular area [17].

In 2018, Kaserekaa et al. offered an intelligent Agent-Based Model to simulate and model evacuees from a building under fire emergency. To assess the suggested model four factors were used, the average time taken to exit (MT), the average potency of the alive people (MP), total deaths (TM), and the total of people alive (TV). When the simulation executed appeared fire spreading, speed, some evacuee people, and other factors could influence the model. Moreover, emotional and physical properties with some other properties such as stress, disability, speed, wind, gender, and age are severely considered by this model and they may considerably affect the decision making of people to evacuate, the author of the proposed model wished to involve these factors due to a fuzzy logic [18]. In 2019, M. Danial et al. developed a simulation model for pedestrian crowd evacuation based on the idea of fuzzy logic techniques, the idea of the KNN algorithm, and some statistical equations. The model defined various speeds for each participant based on different properties such as physical, psychological, and emotional and indicated individuals' evacuation time with their appeared behaviors during the emergency evacuation process. Finally, the model confirmed a combination of various properties, environments, different distribution, and familiarity of the individuals for the environment led to a significant change in the appeared behaviors for the participants and their evacuation efficiency during the emergency evacuation [12].

The table of the authors' contribution to these existing literature works is managed based on table 1 to the contribution of the author(s) in this reference [19]. Table 1 shows the authors' contribution from the perspective of methods they used to build their simulation models, the situation of evacuations, agents who participated in the evacuation process, and appearances that were achieved from the results of the evacuation processes.

**Table 1.** Shows the contribution of authors.

| Authors | Methods | Situations | | Agents | | Appearances |
| --- | --- | --- | --- | --- | --- | --- |
| | | Normal | Emergency | Disable | Not Disable | |
| Nirajan, et al. (2016) | Conduct a model theoretically and mathematically | ✓ | ✓ | | ✓ | focus on emergency controller and evaluating emergency exit signs' locations |
| C. Shuchao, et al. (2018) | An extended multi-grid model | | ✓ | | ✓ | predict the movement of the evacuees, exit choice, and act as a guider |
| Kontou et al. (2018) | A model with Cellular automata (CA) parallel computing tools | | ✓ | ✓ | ✓ | Recording evacuation time |
| Kaserekaa et al. (2018) | An intelligent Agent-Based Model with for factors | | ✓ | | ✓ | Appeared some factors affect the decision making of people to evacuate |
| M. Danial et al. (2019) | Cellular Automata (CA) with fuzzy logic, KNN, and some statistical equations | | ✓ | | ✓ | Records evacuation time and emergency behaviors during the evacuation process |
| This paper | An Integration simulation model | | ✓ | | ✓ | Best design choice among numerous designs |

## 3. Research Method

This section can be divided by subheadings. It provides a concise and precise description of the developed simulation model and the improvement in the developed model.

*3.1. The Developed Simulation Model for Pedestrian Crowd Evacuation*

This section presents the methodology of the most recent simulation model for pedestrian crowd evacuation. It is divided into three main parts; 1) the idea of fuzzy logic that was used to manipulate the individuals' properties via designing various membership functions, and then prepared to be used in defining various individuals' speeds 2) the idea of the kth nearest neighbors (kNN), which was used to make the evacuees find the nearest exit door3) some statistical equations were used to determine the desired speed for each individual through the evacuation process by benefiting from the individuals' properties that were prepared by the idea of fuzzy logic as mentioned.

3.1.1. Idea of Fuzzy Logic in the Developed Model

Inside this developed model, various properties for each individual were collected, for example, physical, emotional, and biological, and then these properties via the idea of fuzzy logic technique were manipulated. This manipulation was used to fuzziness each property and then made the developed model reach a realistic solution. From the fuzziness of each property, the model signified a specific range's qualities spanning to create a linguistic variable, for instance, "disease {very low, low, medium, high, very high}, weight {very slim, slim, heavy, very heavy}, age {adult, very young, young, old, very old}, collaboration {very low, low, medium, high, very high} and shock {very low, low, medium, high, very high}" [12]. Then for each range, a membership function was designed by the model. The membership functions of the age, weight, disease, shock, collaboration are distinctly illustrated in Figure 1a–d [12]. Consequently, these membership functions were utilized in defining the individuals' desired speed.

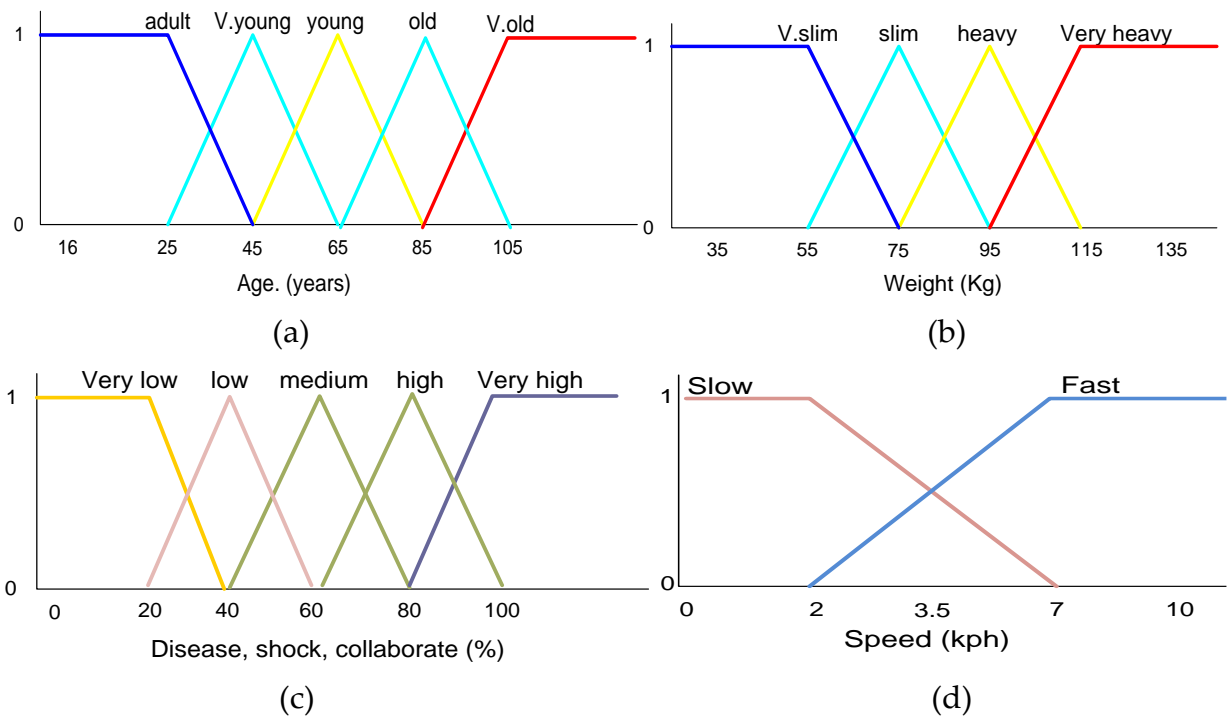

**Figure 1.** Model definition: (**a**) Age membership function, (**b**) Weight membership function, (**c**) Disease, shock, collaborate membership function, (**d**) Speed membership function [12].

3.1.2. Idea of the K-th Nearest Neighbors (KNN) in the Developed Model

K-th Nearest Neighbors (KNN) use the training set and select features in the training set as different dimensions in an area. Having KNN to manage the number of points in the area determines the observation value for each dimension. Hereafter, it measures two points based on their similarity via distance between them. Some suitable metrics indicate that measures, such as Euclidean, Manhattan, and others [20]. Equation (1) in the following is the Euclidian distance used by the developed model to measure two points.

$$\text{dist}((a_1, b_1), (a_2, b_2)) = \sqrt{(a_2 - a_1)^2 + (b_2 - b_1)^2} \tag{1}$$

where $a_2$ and $b_2$ present the coordinate of the exit door location, and $a_1$ and $b_1$ present the coordinate of the pedestrian's location.

Therefore, the algorithm decides to select the new observation's adjacent data points to select the right class among numerous classes [21]. This study works on the improvement in this developed model by adding a new feature to find the best design among numerous designs based on evacuees' evacuation times for each of the designs. Thus, this method was not compared with other methods, such as K means and others, while the developed model already used KNN.

Inside the developed model, participant individuals in the evacuation process were categorized according to the environment's familiarity property through the evacuation area. Their familiarities were determined during the data collection about the participated individuals in the evacuation process. Some participants had no information about the exit door of the evacuation area from the gathered data, whereas some others had. The developed model utilized the K-Nearest Neighbor (KNN) technique to implement the familiar agents by introducing familiar property for agents and checking the distance of each exit Ei inside the classes A, B, and C and then chooses the nearest one to evacuate. Figure 2 shows how evacuees check distances between exit door location and him/herself.

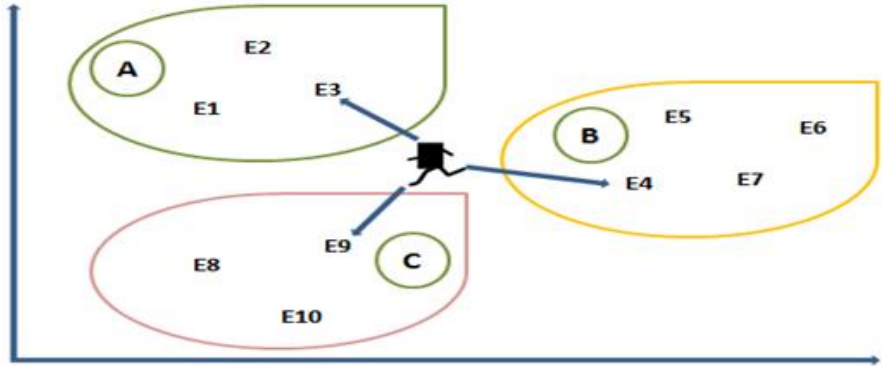

**Figure 2.** Illustrates agent checks the distance of each exit Ei inside the classes A, B, and C based on the K-Nearest Neighbor (KNN) technique and then chooses the nearest one to evacuate.

3.1.3. Statistical Equations Used in the Developed Model

When the participated individuals' properties were gathered and examined, the membership functions created, as mentioned in Section 3.1. These functions were utilized in determining the weighted degree of properties, while the property according to the idea of fuzzy logic consists of two values; lower value and upper value. Figure 1 shows how the weight's degrees of each property values are specified to participate in the speed of the agent.

From Figure 3 appears an agent has age, and the age ranges from 16 to 105 as well as this range is partitioned into some class intervals. Each class interval indicated with a specified name such as 16–25 is adult, 25–45 is very young, 45–65 is young, 65–85 is old, and 85–105 is very old. Based on the fuzzy logic idea, some equations proposed by the author to find the weights of the properties' values, for instance, if age value equals 57 its weight is specified by Equation (2) [2] after specifying the lower value and upper value of the property via using age membership functions. See Figure 3.

$$weightprop = \frac{\sum_{i=1}^{n} wp_i * srd_i}{\sum_{i=1}^{n} wp_i} \tag{2}$$

$wp_i$ denotes the weighted degree for the given property, and $srd_i$ denotes the degree's speed range. This developed model to create heterogeneity inside a single class interval, various forms for Equation (2) were suggested (see Equations (3) and (4)). The agent's degree of weights' property is defined separately by applying Equations (3) and (4) [12].

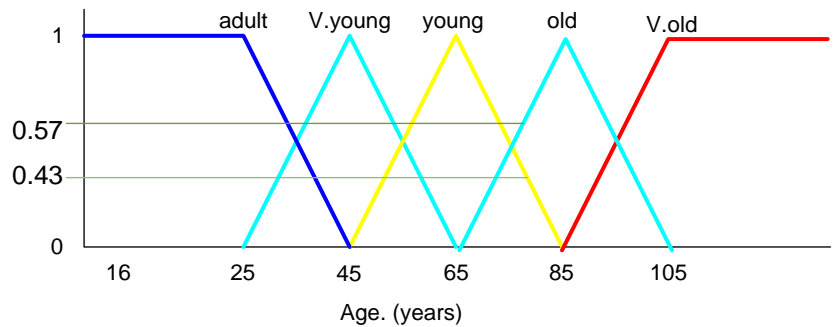

**Figure 3.** Shows how the weights degrees of each property value identified to participate in the speed of the agent [12].

$$weightprop = weightedmean = (lov * minisrd + upv * maxisrd)/(lov + upv) \quad (3)$$

$$weightprop = weightedmean = (upv * minisrd + lov * maxisrd)/(lov + upv) \quad (4)$$

$lov$ denotes lower value and $upv$ denotes upper value, $minisrd$ denotes minimum interval speed range, and $maxisrd$ denotes maximum interval speed range of the mentioned properties. Because these weights participate in specifying the desired speed for an agent the range of the speed must be identified. The speed range is assumed to be a minimum of 2 k/h to a maximum of 7 k/h as shown in Figure 1. Each property has its range as mentioned above and for each property, there is a given value randomly between its ranges or could be chosen by the user of the simulation model. Moreover, this model created several class intervals for the speed range to keep a balance between the property value and the speed. For example, when an agent is 57 years old, this agent according to the designed age membership function 0.57 is young and 0.43 is very young, and its speed range is between 4 k/h to 5 k/h. See Figure 4. The logical reason behind this separation in speed range was older agents are slower than younger agents. Furthermore, Equation (5) (Midvale) [12] was used to find the middle of the chosen class interval by the age membership function.

$$\text{Midvalue} = lov + upv/2 \quad (5)$$

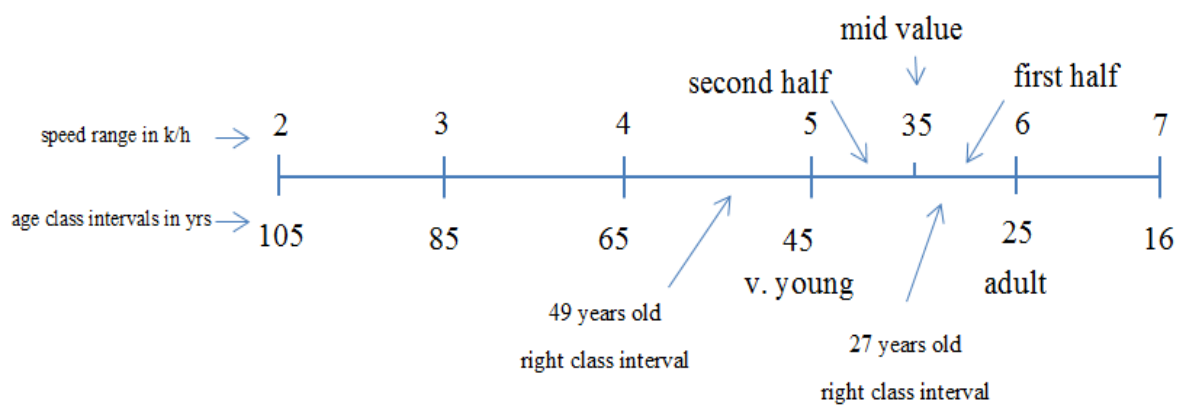

**Figure 4.** How weights of agent's properties are identified [12].

This Midvalue aims to keep diversity in both distinct parts of a chosen class interval and avoid weight redundancy. Consequently, the result of Equation (5) and the value of the property decide on the use of Equation (3) or (4). However, when the given property value was equal to the Midvalue result, there is no difference in Equation (3) or (4). When an agent has more than one property such as age, weight, disease, and so on, the same operations used to find the weight of age would be used to determine the weight for other properties. Finally, Equation (6) [12] would be used to find the desired speed of each agent from the results of Equations (3) and (4) after amalgamation with both emergency factor $em_i$ and gender factor $gen_i$. For readers who are interested to know more about these equations; can look at this paper [12].

$$\text{desiredSpeed} = \left(\frac{\sum_{i=1}^{n} \text{wp}_I}{n}\right) * gen_i * em_i \qquad (6)$$

*3.2. Improvement in the Developed Model*

This section illustrates a proposed method for one of the most recently developed simulation models for pedestrian crowd evacuation. As mentioned in the previous subsection, to improve the model's ability due to finding the best exit door location among various suggested exit door locations through the area of evacuation and selecting a more suitable design among other designs of the evacuation area. Figure 5 presents the proposed method.

> **Step 1:** Specify lower bound and upper bound of the side of the evacuation area, which the designer wants to place the exit door location.
>
> **Step 2:** Identify the number of exit door location, which the designer wants to be tried for the selected side of the evacuation area.
>
> **Step 3:** Get exit door locations randomly between the lower bound and the upper bound as mentioned in step 1.
>
> **Step 4:** Select one of the random exit door locations as mentioned in step 3.
>
> **Step 5:** Start evacuation process for that exit door location from step 4, and then find evacuation time of each individual.
>
> **Step 6:** Find average of individuals' evacuation time which was found by step 5.
>
> **Step 7:** Check if there is still a random exit door location to be tried, if there exists return to step 4, otherwise go to next step.
>
> **Step 8:** Compare averages of the evacuation times for exit door locations.
>
> **Step 9:** Select the exit door locations with the minimum average as the best exit door location to be

**Figure 5.** Shows a proposed method to integrate with the most recent developed simulation models.

This proposed method and integrating with the created simulation modeled to increase the performance of the simulation model due to involving a new capacity to that model via simulating an evacuation area with various exit doors locations with the hope to identify the best location for the exit door among them and also choose the best design for the evacuation area. To confirm this method possibility, various experimentations were done in Section 4 and then discussed.

## 4. Results and Analysis

To approve the possibility of this method and integrating with the developed model, several scenarios have been written, which have been used to execute the model through different experiments. Since the developed model worked on the area of a cafeteria and sufficient data was collected about the individuals that participated in the evacuation process and analyzed, this research also studied the same area and used the same data of the selected model, more interested readers can look at this paper [12]. From there, the written scenarios mentioned above tested for 5 various prominent exit door locations with the same distribution of the individuals through the area of the cafeteria, and the plans could be briefed in these points: (1) only one exit door for each of the student part, employee part and staff part. (2) two exit doors for the student's part, only one exit door for each employee and staff part. (3) two exit doors for the student part, two exit doors for the employee part, and one exit door for the staff part. The above-mentioned points were briefed inside Table 2. For more details, Table 3 provides the importance of some parameters in the final result of evacuation time.

**Table 2.** Shows several scenarios used to execute the model through different experiments.

| Experimentations | Scenarios |
|---|---|
| #A | Each of the employee part, student part, and staff part has one exit door, evacuees were no familiar with the exits |
| #B | Two exit doors for students part, and each of the employee part, and staff part has one exit door, evacuees were no familiar with the exits. |
| #C | Two exit doors for each of the student part, employee part and one exit door for staff part, evacuees were no familiar with the exits |

**Table 3.** Shows the importance of some parameters in the final result of evacuation time.

| No | Parameters | Importance |
|---|---|---|
| 1 | Individuals distribution | Individuals' distribution within a small area of the evacuation leads to more collision among evacuees and decreases the evacuees' speeds the evacuees' speed; thus, evacuation time will increase. On the other hand, when they distribute within a large area, the collision will decrease among evacuees. The evacuees can move toward the exits at their desired speeds; therefore, evacuation time will decrease. An individual takes more time to evacuate when the individual from the distribution is far from the exit door. |
| 2 | Number of each part exit doors within the area of evacuation | The evacuation time is minimized with an increasing number of exit doors for each part within the evacuation area. However, the individuals who choose the wrong exit door to evacuate without having familiarity with the evacuation area may take a long time. |
| 3 | Familiarity | Individuals' familiarity with the area evacuation will help to evacuate in a shorter time from the evacuation area. However, this familiarity causes congestion in the way to the exit door. At this time, the evacuation time will increase. |

All scenarios mentioned inside Table 2 has been tested for 20 evacuees, and each scenario tried for 20 none-familiar evacuees. All the evacuees involved in different scenarios had the same attributes, such as evacuees' ages between 20–57, and evacuees' weights were between 57–102-kg. Moreover, the developed model defined the evacuees' desired speed based on their properties, as mentioned in Section 3.1.3.

*4.1. Result and Experimentation #A*

Inside the model, the evacuation area was designed and managed according to scenario number #A mentioned inside Table 2. The design was tested, as shown in Figure 6, and the results of 20 nonfamiliar evacuees for 5 exit door locations were briefed in Table 4.

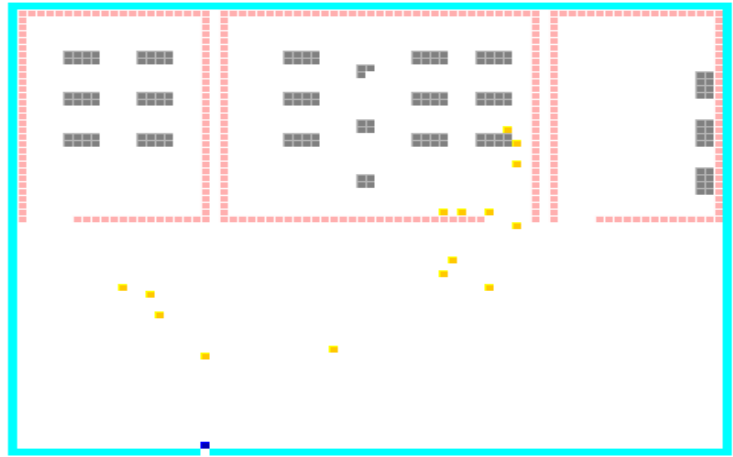

**Figure 6.** Shows the area of evacuation with one exit door for each part of the evacuation area.

**Table 4.** Result of scenario #A for 20 nonfamiliar evacuees for 5 exit door locations.

|  | Exit 1 At Location (70,47) | Exit 2 At Location (70,57) | Exit 3 At Location (70,73) | Exit 4 At Location (70,83) | Exit 5 At Location (70,35) |
|---|---|---|---|---|---|
| **Individuals** | | | **Durations** | | |
| 1 | 0:18:225 | 0:18:158 | 0:19:346 | 0:18:162 | 0:18:203 |
| 2 | 0:11:834 | 0:14:579 | 0:19:330 | 0:22:362 | 0:11:763 |
| 3 | 0:13:990 | 0:12:905 | 0:13:732 | 0:14:260 | 0:16:495 |
| 4 | 0:14:310 | 0:14:752 | 0:15:387 | 0:15:27 | 0:15:851 |
| 5 | 0:13:903 | 0:16:437 | 0:20:197 | 0:26:246 | 0:13:474 |
| 6 | 0:9:289 | 0:9:114 | 0:9:551 | 0:9:686 | 0:10:205 |
| 7 | 0:15:902 | 0:14:552 | 0:15:352 | 0:15:777 | 0:14:585 |
| 8 | 0:13:946 | 0:14:44 | 0:14:819 | 0:15:3 | 0:15:621 |
| 9 | 0:15:469 | 0:15:333 | 0:15:595 | 0:15:637 | 0:15:894 |
| 10 | 0:17:786 | 0:18:444 | 0:16:792 | 0:16:730 | 0:18:307 |
| 11 | 0:10:766 | 0:11:427 | 0:17:157 | 0:19:650 | 0:10:881 |
| 12 | 0:13:192 | 0:15:532 | 0:21:815 | 0:25:356 | 0:13:26 |
| 13 | 0:8:77 | 0:8:969 | 0:13:285 | 0:14:679 | 0:8:496 |
| 14 | 0:19:680 | 0:19:677 | 0:20:238 | 0:20:59 | 0:20:669 |
| 15 | 0:18:526 | 0:18:438 | 0:17:846 | 0:18:839 | 0:19:328 |
| 16 | 0:17:319 | 0:18:365 | 0:17:852 | 0:18:526 | 0:19:231 |
| 17 | 0:15:454 | 0:14:762 | 0:15:431 | 0:14:798 | 0:15:670 |
| 18 | 0:18:773 | 0:18:960 | 0:18:620 | 0:19:264 | 0:19:235 |
| 19 | 0:21:643 | 0:21:811 | 0:21:688 | 0:21:856 | 0:22:118 |

| 20 | 0:22:740 | 0:22:221 | 0:22:534 | 0:22:668 | 0:22:549 |
| --- | --- | --- | --- | --- | --- |
| Average Durations | 15.541 | 15.924 | 17.328 | 18.229 | 16.08 |

*4.2. Result and Experimentation #B*

Inside the model, the evacuation area was designed and managed according to scenario number #B mentioned inside Table 2. The design was tested, as shown in Figure 7, and the results of 20 nonfamiliar evacuees for 5 exit door locations were briefed in Table 5.

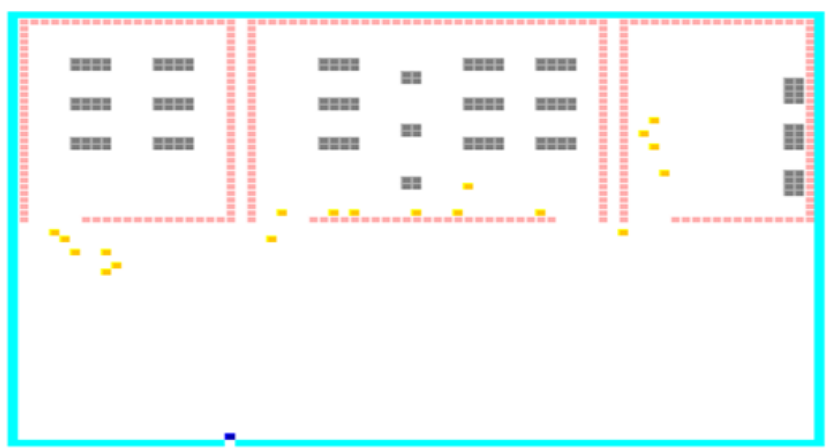

**Figure 7.** Shows the area of evacuation with two exit doors for the student part of the cafeteria.

**Table 5.** Result of scenario #B for 20 nonfamiliar evacuees for 5 exit door locations.

|  | Exit 1 At Location (70,47) | Exit 2 At Location (70,57) | Exit 3 At Location (70,73) | Exit 4 At Location (70,83) | Exit 5 At Location (70,35) |
| --- | --- | --- | --- | --- | --- |
| **Individuals** | | | Durations | | |
| 1 | 0:18:753 | 0:18:661 | 0:18:613 | 0:19:78 | 0:17:41 |
| 2 | 0:12:606 | 0:14:111 | 0:19:264 | 0:22:504 | 0:12:354 |
| 3 | 0:13:462 | 0:13:362 | 0:13:661 | 0:13:751 | 0:16:409 |
| 4 | 0:10:578 | 0:15:691 | 0:9:974 | 0:15:464 | 0:15:770 |
| 5 | 0:12:967 | 0:17:264 | 0:21:40 | 0:23:75 | 0:13:135 |
| 6 | 0:9:384 | 0:14:676 | 0:9:567 | 0:9:907 | 0:10:183 |
| 7 | 0:14:784 | 0:20:895 | 0:21:692 | 0:25:475 | 0:14:575 |
| 8 | 0:14:233 | 0:15:794 | 0:14:105 | 0:18:645 | 0:14:345 |
| 9 | 0:16:361 | 0:14:649 | 0:15:314 | 0:18:279 | 0:15:859 |
| 10 | 0:17:858 | 0:18:396 | 0:14:854 | 0:17:200 | 0:13:450 |
| 11 | 0:11:603 | 0:12:982 | 0:17:131 | 0:20:522 | 0:10:110 |
| 12 | 0:14:300 | 0:16:321 | 0:22:623 | 0:24:119 | 0:14:481 |
| 13 | 0:8:547 | 0:9:425 | 0:12:264 | 0:14:723 | 0:7:918 |
| 14 | 0:20:462 | 0:14:224 | 0:16:56 | 0:19:720 | 0:13:792 |
| 15 | 0:13:569 | 0:18:425 | 0:18:809 | 0:18:183 | 0:18:763 |
| 16 | 0:17:383 | 0:17:615 | 0:18:314 | 0:18:615 | 0:19:220 |
| 17 | 0:14:677 | 0:14:525 | 0:14:273 | 0:14:498 | 0:16:816 |
| 18 | 0:17:787 | 0:17:345 | 0:17:362 | 0:17:665 | 0:19:941 |
| 19 | 0:18:597 | 0:18:829 | 0:18:526 | 0:18:842 | 0:21:487 |
| 20 | 0:21:237 | 0:21:482 | 0:21:551 | 0:21:459 | 0:24:884 |
| Average Durations | 14.957 | 16.233 | 16.749 | 18.229 | 15.526 |

*4.3. Result and Experimentation #C*

Inside the model, the evacuation area was designed and managed according to scenario number #C mentioned in Table 2. The design was tested, as shown in Figure 8, and the results of 20 nonfamiliar evacuees for 5 exit door locations were briefed in Table 6.

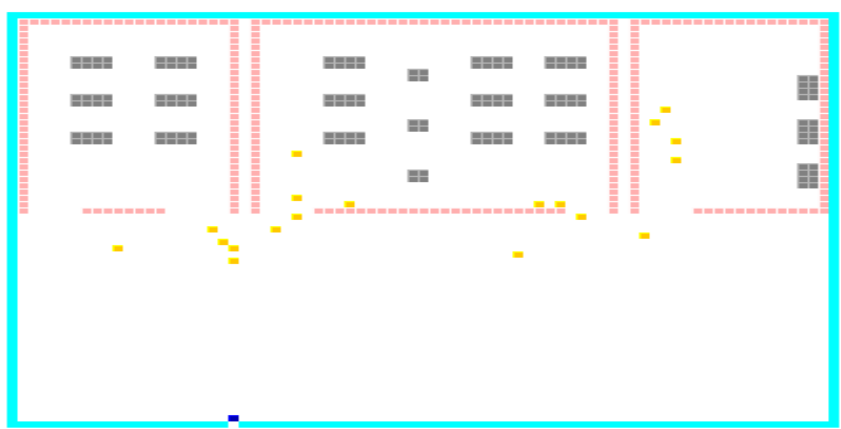

**Figure 8.** Shows the area of evacuation with two exit doors for the student part of the cafeteria.

**Table 6.** Result of scenario #C for 20 nonfamiliar evacuees for 5 exit door locations.

| | Exit 1 At Location (70,47) | Exit 2 At Location (70,57) | Exit 3 At Location (70,73) | Exit 4 At Location (70,83) | Exit 5 At Location (70,35) |
|---|---|---|---|---|---|
| **Individuals** | | | **Durations** | | |
| 1 | 0:17:694 | 0:18:230 | 0:18:618 | 0:19:5 | 0:17:437 |
| 2 | 0:14:820 | 0:14:669 | 0:16:970 | 0:23:706 | 0:14:939 |
| 3 | 0:13:755 | 0:13:632 | 0:14:32 | 0:14:375 | 0:16:418 |
| 4 | 0:15:68 | 0:10:235 | 0:10:2 | 0:12:559 | 0:14:261 |
| 5 | 0:14:2 | 0:12:882 | 0:16:171 | 0:19:442 | 0:13:492 |
| 6 | 0:9:421 | 0:9:515 | 0:15:487 | 0:16:449 | 0:14:126 |
| 7 | 0:15:905 | 0:15:63 | 0:19:834 | 0:25:400 | 0:19:386 |
| 8 | 0:14:871 | 0:14:766 | 0:15:729 | 0:14:806 | 0:15:654 |
| 9 | 0:16:387 | 0:15:307 | 0:15:994 | 0:18:213 | 0:15:554 |
| 10 | 0:14:968 | 0:13:871 | 0:18:92 | 0:17:464 | 0:14:119 |
| 11 | 0:11:84 | 0:12:1 | 0:12:801 | 0:15:343 | 0:10:892 |
| 12 | 0:14:671 | 0:14:933 | 0:18:79 | 0:21:573 | 0:13:78 |
| 13 | 0:8:113 | 0:9:43 | 0:10:55 | 0:16:31 | 0:8:739 |
| 14 | 0:15:74 | 0:19:632 | 0:19:909 | 0:20:736 | 0:14:188 |
| 15 | 0:18:35 | 0:17:903 | 0:18:720 | 0:18:848 | 0:17:543 |
| 16 | 0:17:412 | 0:16:937 | 0:18:222 | 0:18:669 | 0:16:903 |
| 17 | 0:14:368 | 0:14:545 | 0:14:193 | 0:14:566 | 0:16:880 |
| 18 | 0:17:793 | 0:17:353 | 0:17:295 | 0:17:720 | 0:20:607 |
| 19 | 0:18:609 | 0:18:841 | 0:18:430 | 0:18:882 | 0:21:462 |
| 20 | 0:21:77 | 0:21:139 | 0:21:460 | 0:21:548 | 0:24:899 |
| Average Durations | 15.156 | 15.024 | 16.504 | 18.266 | 16.028 |

From this integrated model's experimentations' results, it appeared the mentioned proposed method in Section 3.2 worked properly. The method made the developed model significantly improve in finding the evacuees' evacuation times' averages for various exit door locations and then used them to select the best exit door location among them. Moreover, the written scenarios as mentioned above made the model create different designs for the area of evacuation to be tested by the improved model. Consequently, the proposed method worked as a new capability of the developed model after it compared the evacuees' evacuation times' average for the suggested exit doors' locations. It indicated the best exit door location among many others via the selection of minimum average durations that all evacuees took to evacuate from the evacuation area's exit door locations. The Green color inside Tables 4–6 shows the minimum average duration belonged to the best exit door location.

When change occurs in the evacuation area's design, it changes the evacuees' evacuation time within the evacuation process. Thus, it changes decision making to select the best exit door location for evacuation, among others. For instance, in Figures 6–8, an evacuation area contains three different parts. As mentioned in the developed model in paper [12] first part is the employees' part, the second part is the students and the third part is the staffs' part. As shown in Figure 6, each part has one exit door. From the results shown in Table 4, the best exit door location was the (70,47) with an average duration of 15 s and 541 milliseconds. Moreover, as shown in Figure 7, the students' part has two exit doors while the employees' part and the staff's part has one exit door. From the results presented in Table 5, the evacuation duration was significantly improved. However, the best exit door location remains the same (70,47) with 14 s and 957 milliseconds. Furthermore, as shown in Figure 8, the employees' part and students' part have two exit doors while the staff's part has one exit door. From the results obtained in Table 6, the evacuation duration was slightly increased and the best exit door location changed to (70,57) with a duration of 15 s and 024 milliseconds. Finally, the designer can choose the best design for the evacuees from the three tested designs. The design shown in Figure 7 was safer than other designs shown in Figures 6 and 8 while it takes only 14 s and 957 milliseconds for evacuees to evacuate from the evacuation area. See Figure 9.

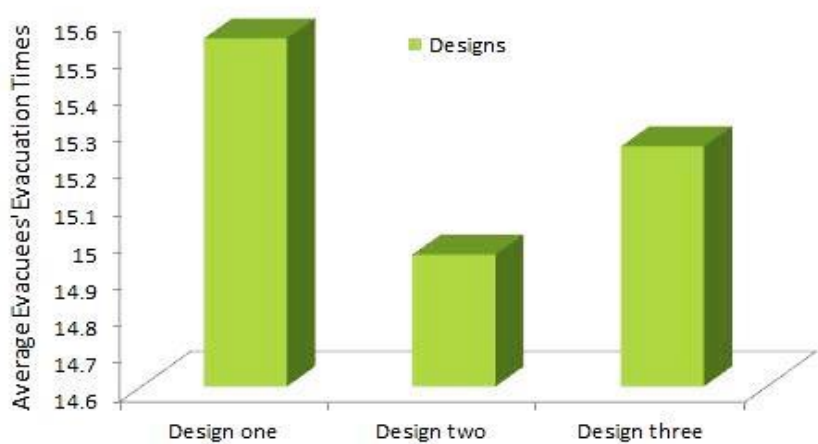

**Figure 9.** Evacuation times for three different designs of the evacuation areas.

## 5. Conclusions

Even though identifying evacuation time and appearing behaviors during emergency evacuation for evacuees is vital, it is still crucial to build methods for these models to simulate various circumstances to enlarge the model's capabilities carefully. From there, in this study, a new method was proposed and integrated with one of the most recent simulation models for pedestrian crowd evacuation published in late 2019. This integration has been made to increase the ability of the developed model and make the model simulate affect the different exit door locations on pedestrians' evacuation time and choose the best exit door location among them. This method was dependent on the developed model's output. In contrast, it used the evacuees' evacuation times to specify the best exit door location, among others.

Furthermore, to confirm this proposed method's ability, several scenarios are written and tried by the integrated model. From the results appeared the developed model worked more capable than before, whereas, with the proposed method, the average of the evacuees' evacuation times could be found and then use to determine the best exit door location among many others. From there, the designers made a decision on which design is the best for an evacuation area in terms of safety. Finally, it is recommended as future work to focus on adding a new feature to the proposed method to determine which exit door had more congestion and collision between the evacuees and determine how the congestion and collision influenced the evacuation process.

**Author Contributions:** Conceptualization, D.A.M., and T.A.R.; methodology, D.A.M., and T.A.R.; software, D.A.M.; validation, A.A., N.B., and P.F.; formal analysis, D.A.M., and T.A.R.; investigation, A.A., N.B., P.F.; resources, T.A.R.; data curation, D.A.M.; writing—original draft preparation, D.A.M.; writing—review and editing, T.A.R.; visualization, P.F., M.M., I.B.; supervision, T.A.R. All authors have read and agreed to the published version of the manuscript.

**Funding:** This research received no external funding.

**Acknowledgments:** The authors would like to thank the universities involved in providing facilities and equipment for this review work.

**Conflicts of Interest:** During this research, there were no conflicts of interest.